%% file: SIR_Graph_Growth_v3.tex
\documentclass[a4paper]{amsart}  
\usepackage[english]{babel}
\usepackage[utf8]{inputenc}
\usepackage{algorithmic}
\usepackage{algorithm}
\usepackage{amsfonts}
\usepackage{amsmath}
\usepackage{amssymb}
\usepackage{amsthm}
\usepackage{enumerate}
\usepackage{epsfig}
\usepackage{graphicx}
\usepackage{amsaddr}
\usepackage{subfigure}
\usepackage{url}

\input{CsdMacros}

\begin{document}

\title[An SIR Graph Growth Model for Epidemics]{An SIR Graph Growth Model for the Epidemics \\of Communicable Diseases}

\author{Charanpal Dhanjal \and St\'ephan Cl\'{e}men\c{c}on}
\address[Charanpal Dhanjal]{T\'{e}l\'{e}com ParisTech, 46 rue Barrault, 75634 Paris Cedex 13, France}
\email[Charanpal Dhanjal]{\{charanpal.dhanjal, stephan.clemencon\}@telecom-paristech.fr}

\date{\today}

\begin{abstract} 
It is the main purpose of this paper to introduce a graph-valued stochastic process in order to model the spread of a communicable infectious disease. The major novelty of the SIR model we promote lies in the fact that the social network on which the epidemics is taking place is not specified in advance but evolves through time, accounting for the temporal evolution of the interactions involving infective individuals. Without assuming the existence of a fixed underlying network model, the stochastic process introduced describes, in a flexible and realistic manner, epidemic spread in non-uniformly mixing and possibly heterogeneous populations. It is shown how to fit such a (parametrised) model by means of \textit{Approximate Bayesian Computation} methods based on graph-valued statistics. The concepts and statistical methods described in this paper are finally applied to a real epidemic dataset, related to the spread of HIV in Cuba in presence of a contact tracing system, which permits one to reconstruct partly the evolution of the graph of sexual partners diagnosed HIV positive between 1986 and 2006.
\end{abstract}


\maketitle

\section{Introduction}
\label{sec:intro}

Although a wide collection of variants of the so-termed \textit{standard stochastic SIR model} introduced in \cite{kermackmckendrick,bartlett} have been proposed to describe ever more realistically the situations encountered in practice (presence of control strategies, possibility of reinfection, \textit{etc.}), modelling of the spread of transmissible infectious diseases is still the subject of a good deal of attention in mathematical epidemiology. Most of the models proposed in the literature stratify the population under study into \textit{compartments} that describe all possible serological statuses related to the disease and formulate assumptions (of probabilistic nature) that governs the transfer of individuals from one compartment to another (refer to \cite{AndersonBritton,mode} for a review of classical stochastic models in the epidemiology context). The unrealistic hypothesis of a \textit{homogeneously mixing population} (roughly saying that a given infectious individual may transmit the disease to any susceptible individual at the same rate) has been progressively abandoned in the large population framework. It does not account for the possibility of repeated contacts between pairs of individuals that know each other. Graph structures, representing the underlying social network structure, are more and more frequently incorporated to epidemic models. References are much too numerous to be listed exhaustively, see \cite{barthelemybarratpastorsatorrasvespignani,durrett,newman03graphs,volz,eubank2004modelling} for instance. The general idea is to specify a graph structure in advance and let the infectious disease spread on it (the term \textit{percolation} is then used in general). Amongst the most widely used models  are the Erd\"{o}s-R\'enyi graphs (abusively referred to as ``random graphs''; see \cite{erdos1959random,erdos1960random} as well as \cite{durrett} and the references therein for epidemiological processes taking place on such graphs), small-world graphs (see \cite{ ballmollisonscaliatomba,barbourreinert,watts98smallWorld} and also \cite{moore00epidemics}) or configuration models (see \cite{bollobas2001,molloyreed,vanderhofstad} as well as \cite{ballneal,decreusefonddhersinmoyaltran,volz}), depending on the properties of the networks under study (connectivity, transitivity, mixing patterns, \textit{etc.}). A good review of the subject is presented in \cite{keeling2005networks}. The concept of social network also permits one to model efficiently the impact of active detection strategies of the \textit{contact tracing} (CT) type, widely used to control the spread of sexually transmissible diseases. The graph structure is involved explicitly: individuals diagnosed positive are asked to name those with whom they have had possibly infectious contacts, the latter, when identified,  are then offered a medical examination and possibly a cure in case of infection.

\par However, when considering long-lasting epidemics or endemic situations, the assumption that the social network on which the epidemics occurs remains fixed becomes highly questionable. In the HIV case for instance, the period during which two individuals are sexual partners may be very short compared to the duration of the epidemic. In contrast,  this paper proposes a graph-valued epidemiological process relying on an \textit{individual-based dynamics}, without specifying \textit{a priori} an underlying graph structure or holding the number of contacts (\textit{i.e.} the degree) of the individuals in the population under study fixed. We generate a \textit{time-evolving graph}, where the occurrence of an infectious contact between two individuals depends on their own characteristics and on their recent connections both at the same time. This way, new contacts are established according to a stochastic procedure that echoes the concept of \textit{preferential attachment} in graph-mining, see \cite{ newman03graphs}, and thus accounts for real-life situations. As mentioned above, this approach also facilitates the modelling of the effects of contact tracing. 

Beyond the description of the probabilistic model of network evolution for the spread of communicable infectious diseases we propose, issues related to the statistical calibration of the parameters driving the epidemiological process are also tackled, from the angle of \textit{Approximate Bayesian Computation} (ABC). Parameter estimation for SIR models is usually a difficult task, mainly because the epidemiological process is in general partially observed: a significant part of the information about infectious contacts can be missing (time, frequency, identity of the transmitters, \textit{etc.}). Hence, in most cases, the model likelihood is of the form of an integral whose computation is numerically infeasible, even approximately by means of Markov Chain Monte Carlo (MCMC) methods (see \cite{oneillroberts,oneill,cauchemcarrat}) in certain situations. Indeed, MCMC techniques can be computationally prohibitive for high-dimensional missing observations (the number of unobserved events can be even unknown and arbitrarily large in certain cases) and take several days on parallelised systems, see \cite{chisstersinghferguson}. Originally proposed for making inference in population genetics \cite{pritchardetal,pritchard02,beaumont02genetics}, ABC offers a promising alternative for statistical inference in epidemiological models \cite{blum10biostat,mckinley09ijb,walker09sma} (see also \cite{blum10jasa} for the theoretical evaluation of approximation errors). The ABC approach is not based on the likelihood function but relies on numerical simulations and comparisons between simulated and observed summary statistics. Here, the major novelty arises from the nature of the summary statistics considered for the implementation of the ABC procedure, the latter reflecting the main observable features of the evolution of the social network on which the epidemic takes place. The relevance of our approach is first supported by numerical experiments based on toy models. It is next applied to real epidemiological data, concerning the HIV epidemic in Cuba: the data repository at our disposal gathers information about 5389 individuals diagnosed as HIV positive between 1986 and 2006, who are linked together by $4073$ declared sexual contacts. As will be shown, the modelling promoted here permits one to reproduce accurately a variety of properties of the incidence process, related to the evolution of the sexual contact network of the individuals diagnosed HIV positive. The Cuban HIV epidemic will serve as a running example throughout the article, the data aforementioned being the main source of inspiration of the present work.

\par The article is organized as follows. In Section \ref{sec:epidemicModelling}, we describe the epidemic model we propose at length and in full generality. For illustration purposes, we also show how it applies to the Cuban HIV epidemic, our running example throughout the article (see \cite{dhanjal2011evolution,clemenccon2014statistical}). In Section \ref{sec:inference}, we show that estimation can be carried out by means of ABC methods in practice, to overcome difficulties caused by missing observations in epidemiological data. Finally, the application of the concepts and statistical techniques developed here to the Cuban HIV data is presented in Section \ref{sec:numerical}, together with an in-depth discussion of the results obtained. Technical details are deferred to the Appendix.

\section{Spread of a Transmissible Disease on a Time-evolving Network}\label{sec:epidemicModelling}

We first present a general model of epidemics spreading on an evolving graph, where individuals make contact according to their characteristics and past relations. The model bears similarities with preferential attachment models but also with the SIR models on configuration models, proposed in \cite{ballneal,decreusefonddhersinmoyaltran}, which considered non-structured populations and a fixed degree for each individual. This modelling is applied to the Cuban HIV/AIDS epidemic, for which explicit forms for the rate functionals and parameters are proposed. Here and throughout, we will denote by $\mathbb{I}\{\mathcal{E}\}$ the indicator function of any event $\mathcal{E}$, a matrix with a bold uppercase letter e.g. $\Am$, and a vector with a bold lowercase letter e.g. $\bv$.  

\subsection{A General Model of an Evolving Epidemic}

Here we propose a very general and elementary stochastic model for epidemics in which individuals are explicitly represented as well as the contact between individuals over which disease is transmitted. For the sake of simplicity, it is assumed that the population is closed to immigration and emigration, and births and deaths are not considered. The approach subsequently developed could be straightforwardly extended to a more general framework, accounting for demographic features. The type of contact modelled corresponds to that over which the disease is spread and in the case of the HIV epidemic, these are sexual contacts. Taking arbitrarily $0$ as time origin, the model describes the evolution of a marked graph-valued process $G = (G_t)_{t\geq 0}=(V, E_t, \Xm_t)_{t\geq0}$ with vertices $V = \{v_1, \ldots, v_{M}\}$, $M$ being the fixed population size and undirected edges $E_t \in V \times V$ describing pairs of individuals in contact at time $t\geq 0$. We write $v_i\sim_t v_j$ if there is an edge between the nodes $v_i$ and $v_j$ at time $t$. Furthermore, the matrix $\Xm_t=(\xv_{i}^t)_{1\leq i\leq M}$ has rows composed of descriptor vectors $\xv_{i}^t$ assigned to the $i$th individual which take values in a set $\mathcal{X}\subset \mathbb{R}^d$ say, and will be involved in the rates of various events within the model. The label $\xv_i^t$ describes in particular the serological status of the individual/vertex: susceptible (S), infective (I) or removed/detected (R). It can also encapsulate any covariate relevant for the modelling of the transmission and detection mechanisms, such as its current degree within the graph (\textit{i.e.} the number of individuals whom she/he is currently in contact with), formally $d_t(v_i)=\sum_{j\neq i}\mathbb{I}\{v_i\sim_t v_j\}$ for $i\in\{1,\;\ldots,\; M\}$. In the HIV situation, it can store the date of birth, gender, sexual orientation (heterosexual or bisexual), location, \textit{etc.} The sets of individuals in the serological states $S$, $I$ and $R$ at time $t\geq 0$ are respectively denoted by $\mathcal{S}_t$, $\mathcal{I}_t$ and $\mathcal{R}_t$. We also denote by $|\mathcal{S}_t|$, $|\mathcal{I}_t|$ and $|\mathcal{R}_t|$ the sizes of these classes. The population evolves through the following phenomena.
\medskip

\begin{itemize}
\item {\bf Contact:} Individuals can establish contacts between each other, which can result in a disease transmission. We assume that once an individual is detected, she/he no longer makes contacts. Contacts between an individual with characteristics $\xv\in \mathcal{X}$ and an individual described by $\xv'\in \mathcal{X}$ occur with rate $c(\xv,\xv')$.
\item {\bf Infection:} When an infected individual with characteristics $\xv$ makes contact with a susceptible individual described by $\xv'$, the latter becomes infected with probability $f(\xv,\xv')\in [0,1]$.
\item {\bf Detection:} Each infective, with characteristics $\xv$ can be detected by \textit{spontaneous detection}, with rate $\gamma(\xv)$, or independently through \textit{contact tracing}: each detected individual described by $\xv'$ with whom she/he had contacts exerts on her/him a detection pressure of rate $\beta(\xv',\xv)$.
\end{itemize}
\medskip

If explicit forms for the functionals $c,\; f,\; \gamma$ and $\beta$ are available, one can write stochastic differential equations to describe mathematically the evolution of the population, using Poisson point processes. This, together with acceptance-rejection algorithms to handle rates which depend on time and random group sizes, allows exact simulations to be performed. Technical details for this model, including pseudo-code, are given in Appendix \ref{sec:alg}. 

\subsection{Running Example: The Cuban HIV Epidemic}\label{sec:HIVEpidemicModel}

As an application, we specialise the above model for the Cuban HIV epidemic. Here, the time-evolving mark $\xv_{i}^t$ describes the current and past connectivity properties of the individual/node indexed by $i=1, \ldots, M$. Events are simulated from Poisson point processes, with an acceptance-rejection procedure when rates depend on time.

\subsubsection{Contacts} 

Sexual contact is the main vector of transmission of HIV, with only a negligible number of transmissions occurring through blood transfusion, shared needle use and birth. Each individual $v_i$ makes contacts with persons of the appropriate sex at rate $\lambda(\xv_{i}^t)$ which depends on its characteristics (sexual orientation, gender, serological status). When a contact is made, the partner is either a previous contact or a new one. A new partner is chosen using a procedure very close to preferential attachment (see \cite{newmanstrogatzwatts2}). 

For this purpose, we associate with each individual an initial number of former sexual partners. The initial degrees of the vertices, $d_h(v_i)=\sum_{j\neq i}\mathbb{I}\{v_i\sim_0 v_j\}$ for $i\in\{1,\;\ldots,\; M\}$, can for instance be drawn from a power law distribution. For the HIV/AIDS epidemic in Cuba, the database records only the connections which have occurred during the last two years before detection: some previous edges can be assessed through earlier detections of contacts, but the list of contacts is partially observed. A complication of choosing vertices according to their degrees is that we have an incomplete graph of the sexual contact network since we only model contact between infected and non-removed individuals and it follows that the degrees are only known for an individual for the period after infection has occurred.

For a vertex $v_i$, the previous partner $v_j$ is chosen with probability $\alpha$ otherwise a new partner $v_k$ is chosen. Let $\mathcal{C}_{ij}$ denote the set of non-removed individuals which are sexually compatible with $v_i$ excluding the last partner $v_j$. A possible partner $v_k \in \mathcal{C}_{ij}$ is chosen proportionally to its degree, \textit{i.e.} with probability
\begin{displaymath} 
\frac{(1-\tau) d(v_k)+ \tau d_h(v_k)}{\sum_{v_\ell \in \mathcal{C}_{ij}}(1-\tau) d(v_\ell)+\tau d_h(v_\ell)},
\end{displaymath}
where $d(v_k)$ is the observed degree of the individual, $d_h(v_k)$ is the initial degree and $\tau \in [0,1]$ is a user-defined trade off between the initial degree of a vertex and its observed degree. If we choose $\tau=0$ then we have a preferential attachment that depends only on the observed degree $d(v)$. Notice that the choice of the contact among individuals with compatible sexual orientation depends only on the degrees and not on the covariates for the sake of simplicity. The approach can be straightforwardly extended to more sophisticated models. This model mimics the preferential attachment actually observed in the social network built from the Cuban database in which individuals with higher degrees are more likely to form sexual contact. 

\subsubsection{Infections} 

On the occasion of a sexual contact, an infected individual can infect her/his partner according to a probability $f(\xv^t_i,\xv^t_j)$ which depends on the properties of the pair of individuals. When modelling the HIV Cuban epidemic, only sexual orientation is taken into account and $\sigma$ denotes the infection probability for women infecting men, men infecting women and bisexual men infecting men. It is assumed that women cannot infect each other.

\subsubsection{Detections} 

Random detections are assumed to occur at a constant rate $\gamma$ for each infected individual. Independently, an infectious individual $v_i$ at time $t$ can be removed because one of her/his contacts $v_j$ has been previously detected, at time $t(v_j)\leq t$. We assume that after a person is detected, her/his history of contacts is searched for a period of between 6 months and 2 years after the date of detection. Thus, the detection rate of an infected individual $v_j$ is given by:
\begin{displaymath}
\gamma + \beta \sum_{j:\; \exists s\leq t \textit{ s.t. } v_j\sim_s v_i} \mathbb{I}\{t(v_j)\in [t-\eta_1,\; t-\eta_2]\},
\end{displaymath} 
where $\eta_1 = 24 \times 30$ and $\eta_2 = 6\times 30$ represent the upper and lower time limits on the contact tracing period. 

\subsection{Computational Issues}

An apparent difficulty of the above model is the high computational complexity of simulating sexual contacts in a large network. To mitigate this effect, we simulate contacts only involving at least one infected individual, and later we show that the non-simulated contacts are effectively modelled using the hidden degree distribution in order to exhibit properties commonly observed in real social networks. In reality, an infected individual often makes contact with a few other non-removed individuals. Denote the upper bound on the number of contacts per infective as $\kappa$, then the complexity of computing all contact rates at time $t$ is then $\mathcal{O}(|\mathcal{I}_t| \kappa)$ in which $\mathcal{I}_t$ is often small compared to $M$. Detection rates can be computed at a cost proportional to the number of infected individuals. In practice, a simulated epidemic covering a period of several years and involving 1000s of vertices can be simulated rapidly, a fact which is important for the ABC parameter estimation described in the next section. 

\section{Statistical Inference: The ABC Approach}\label{sec:inference}

We now explain the principles of ABC estimation in the context of the model described in Section \ref{sec:HIVEpidemicModel}.

\subsection{Data and Statistics}
Beyond the very general and flexible generative epidemics model described in the previous section, the major purpose of this article is to develop computationally efficient tools for statistical inference based on information at the individual level. In the contact tracing situation, a contact database is built in a sequential manner each time a detection occurs, where detected individuals are listed together with their attributes and declared contacts. It is thus possible to rebuild the evolution of the sexual contact network among individuals who have been diagnosed as infected. Define $\mathcal{G}=(\mathcal{G}_t)_{t\geq 0}$ as the corresponding marked graph-valued process, which will be referred to as the \textit{observable network} in the sequel. At time $t\geq 0$, this graph counts $\mathcal{R}_t = \{v_i :  t(v_i) \leq t\}$ vertices among those of $G_t$, namely $\mathcal{V}_t= \mathcal{R}_t$. We may thus write $\mathcal{G}_t=(\mathcal{V}_t,\mathcal{E}_t, \mathbf{X}_t)$ with $\mathbf{X}_t=(\mathbf{x}_{i}^t)_{v_i \in \mathcal{R}_t}$, where $\mathcal{E}_t$ is a set of edges between vertices in $\mathcal{R}_t$ . Hence, at any time $T\geq 0$, estimation must rely on the path $(\mathcal{G}_t)_{t\in [0,T]}$, or on certain summary properties, \textit{e.g.} the incidence curve (\textit{i.e.} the mapping $t\in[0,T]\mapsto \mathcal{R}_t$).

\subsection{Approximate Bayesian Computation Using Graph-based Statistics}

As usual in Bayesian statistics \cite{marinrobert}, ABC's principle is to estimate a posterior distribution, given a prior distribution $\pi(\theta)$ of the parameter $\theta$ and data $\textbf{x}$. However, instead of focusing on the posterior density $p(\theta\, |\, \mathbf{x})$, ABC aims at a possibly less informative target density $p(\theta\, |\, S(\mathbf{x})=s_{obs})\propto {\textrm Pr}(s_ {obs} |\theta)\pi(\theta)$ where $S$ is a summary statistic that takes its values in a normed space, and $s_{obs}$ denotes the observed summary statistic. The summary statistic $S$ can be a $d$-dimensional vector or an infinite-dimensional variable such as a $L^1$ function for example. Of course, if $S$ is sufficient, then the two conditional densities are the same. To proceed, ABC simulates random parameters $\theta_i$ in the prior distribution, and from this, a data set $\textbf{x}_i$ is simulated and the associated summary statistic $S(\xv_i)$ is computed, for $i = 1, \ldots, N$. Parameters are accepted if $d(S(\xv_i), s_{obs}) \leq \varepsilon$, given a particular tolerance threshold $\varepsilon$ and similarity measure $d$. 

For the purpose of accelerating the convergence to the target distribution and taking advantage of the exploration of the parameter space, adaptive ABC algorithms where the sampling distribution of the parameter $\theta$ is changed during the procedure have been considered  \cite{robertbeaumontmarincornuet,sissonfantanaka,toni2009approximate}. For clarity, we recall the procedure proposed in \cite{toni2009approximate} as this is the method we use for the parameter estimation of our epidemic model. We start from $N$ parameter values $(\theta^{(1)}_0,\dots ,\theta^{(N)}_0)$ (also called \emph{particles}) drawn from the prior $\pi(\theta)$. The procedure generates for every $t\in \{0,\dots,T\}$ a sample $(\theta^{(1)}_t,\dots ,\theta^{(N)}_t)$ and we modify it recursively until their empirical distribution is close enough to the target distribution. There are three steps to the procedure for each iteration $t$:
\begin{enumerate}
\item For each $t$, let $K_t(.,.)$ be a perturbation kernel and $\varepsilon_t$ be a threshold parameter. To each $\theta^{(i)}_t$ of the current sample, we associate a weight
\begin{displaymath} 
w^i_t=\frac{\pi(\theta^{(i)}_t)}{\sum_{j=1}^N w^{j}_{t-1} K_t(\theta_{t-1}^{(j)},\theta^{(i)}_t)}
\end{displaymath}
for $t>0$ and with $w^i_0=1$.
\item Selection step: we sample $(\theta^{(1)*},\dots,\theta^{(N)*})$ (with replacement) from the empirical distribution of $(\theta^{(1)}_t,\dots ,\theta^{(N)}_t)$ weighted by the $w^i_t$'s. If $t=0$ then the sample is taken from the prior distribution. 
\item Mutation step: we transform the current sample $(\theta^{(1)*},\dots,\theta^{(N)*})$ into a sample $(\theta^{(1)**},\; \dots,\;\theta^{(N)**})$ by using an acceptance-rejection procedure. For this, we simulate $\theta^{(i)**}$ from the distribution $K_t(\theta^{(i)*},.)$ for each $i\in \{1,\dots,N\}$ and associate to it a simulated dataset $\xv_i^{**}$ for which the summary statistics $S(\xv_i^{**})$ is computed. We repeat this simulation of $\theta^{(i)**}$ as long as $d(S(\xv_i^{**}), s_{obs})\geq \varepsilon_t$ or $S(\xv_i^{**})$ is not defined, where $d$ is a similarity measure. When the particle is accepted set $\theta^{(i)}_{t+1}=\theta^{(i)**}$.
\end{enumerate}
The idea is that instead of repeating $N$ independent Monte-Carlo estimations, we consider a whole (interacting) sample. The parameter space is explored by moving each parameter, and selecting the best parameters for the next step. For $N$ sufficiently large, the exploration of the parameter space will not be stuck in areas of low probabilities. In Step 1 and iteration $t$, the denominator $\sum_{j=1}^N w^{j}_{t-1} K_t(\theta_{t-1}^{(j)},\theta^{(i)}_t)$ gives the probability of having $\theta^{(i)}_t$ when we start at a position sampled from the weighted empirical distribution of the $(\theta^{(1)}_{t-1},\dots,\theta^{(N)}_{t-1})$ and perturbed with the transition kernel $K_t(.,.)$. Toni \textit{et al.} propose for instance uniform perturbations: $K_t=\sigma U(-1,1)$ where $U(-1,1)$ is the uniform distribution on $[-1,1]$. Step 2 can be understood as a selection step where we remove particles which are obtained with a probability much lower than if they had been sampled from the prior $\pi(\theta)$. Step 3 is the mutation step, where we explore the parameter space, starting from the ``best'' selected positions. The rejection-acceptance step depends on the threshold $\varepsilon_t$ that may vary. 

\subsection{Graph Matching}

As previously recalled, the ABC approach should be extended so as to find model parameters which produce \textit{observable} networks ``closest'' to the observation $\mathcal{G}$ in average, which leads us to the issue of measuring graph similarity, see \cite{zager2005graph, conte2004thirty}. In the graph similarity problem, we require a ``distance'' metric $d(H, H')$ between two undirected graphs $H = (V, E)$ and $H' = (V', E')$. Notice that $H$ and $H'$ need not have the same number of vertices. One can also consider this problem in terms of an \emph{adjacency matrix} $\Am \in \{0, 1\}^{M \times M}$ which has $\Am_{ij} = 1$ if there is an edge from the $i$-th to the $j$-th vertex otherwise $\Am_{ij} = 0$.  In our case the graph also has a vertex label matrix $\Xm \in \mathbb{R}^{M \times d}$ such that the $i$th row of $\Xm$ is a label for the $i$th vertex. 

We restrict our discussion to distances based on the adjacency matrix as they can often be phrased as (relaxed) continuous optimisations and hence solved using off-the-shelf optimisation algorithms. In particular we consider the approach of \cite{zaslavskiy08path} which seeks to solve 

\begin{displaymath} 
 \min \; (1-\nu) F_0(\Pm) - \nu \langle \Cm, \Pm \rangle_F,
\end{displaymath}
\noindent 
where $F_0(\Pm) = \|\Am - \Pm\Am'\Pm\|^2_F$,   $\| \Am \|_F^2 = \sum_{i,j} \Am_{ij}^2$ is the Frobenius norm, $\nu \in [0, 1]$ is a trade-off between matching labels and graph structure and $\Pm$ is a \emph{permutation matrix}, i.e. a binary matrix in which rows and column sum to 1.  The $ij$th entry of $\Cm$ is the cost of fitness between the $i$th vertex in $G$ and the $j$th in $G'$ (a larger score corresponds to a better matching), and hence the total cost of fitness is given by $\langle \Cm, \Pm \rangle_F$, where $\langle \Um, \Vm \rangle_F = \sum_{i,j} \Um_{ij}\Vm_{ij}$ is the Frobenius inner product. Minimising the above objective is combinatorial in nature and hence the constraint on $\Pm$ is relaxed to be in the set of matrices with row and columns sums equal to one and non-negative entries (known as \emph{doubly stochastic} matrices). In addition, to avoid numerical issues and to ensure that the scales of $\|\Am - \Pm\Am'\Pm\|^2_F$ and $\langle \Cm, \Pm \rangle_F$ are similar, one uses a normalised version of the above objective, 

\begin{displaymath} 
 \min \; \phi(\mathcal{G}, \widehat{\mathcal{G}}) = \min \; (1-\nu) \frac{\|\Am - \Pm\Am'\Pm\|^2_F}{\|\Am\|^2_F + \|\Am'\|^2_F} - \nu \cfrac{\langle \Cm, \Pm \rangle_F}{\|\Cm\|_F},
\end{displaymath}
\noindent 
where the fractional part of the first term is in the range $[0, 1]$ and the second term is greater than zero. 

To detail label matching first let $\hat{\Cm}_{ij} = \|\hat{\xv}_i - \hat{\xv}_j\|$, where $\hat{\xv}_i$ is the $i$th label normalised so that the norm of each column of $\hat{\Xm} = [\hat{\xv}_1, \ldots, \hat{\xv}_M]^T$ is 1. Then the cost of fitness matrix is found with $\Cm_{ij} = (\max(\hat{\Cm}) - \hat{\Cm}_{ij})/(\max(\hat{\Cm}) - \min(\hat{\Cm}))$ for all $i, j$. In this way, entries of $\Cm$ are in the range $[0, 1]$ and similar vertex labels have a small cost of fitness. In the case that the graphs are not the same size, the cost of fitness matrix $\Cm$ of smaller dimension is extended with values $\min(\hat{\Cm})$. For adjacency matrices one pads with values $\xi \geq 0$. 

Going back to the issue of selecting a model on the basis of the temporal evolution of the observable network $(\mathcal{G}_t)_{t\in[0,T]}$, we propose to consider a subdivision $\pi:\; t_0=0<t_1<\ldots<t_K=T$ of the time interval and compute the weighted mean objective: 
\begin{displaymath} 
\Phi_{\pi}(\mathcal{G},\widehat{\mathcal{G}}) = \frac{\sum_{i=0}^K \omega (1- \omega)^{K-i} \phi(\mathcal{G}_{t_i}, \widehat{\mathcal{G}}_{t_i})}{\sum_{i=0}^K \omega (1- \omega)^{K-i}}.
\end{displaymath}
\noindent 
This value is an exponentially weighted average with parameter $\omega \in [0,1]$ where the weight for the $i$th term is $\omega (1- \omega)^{K-i}$. Each term can be negative (although in practice we observed it to be always positive) however this fact is not important when used in the context of ABC. 

\section{Numerical Experiments}\label{sec:numerical}

In this section, we perform a series of simulations using the stochastic epidemic model in conjunction with ABC. There are two primary aims of this work: the first is that we wish to observe how effectively ABC predicts parameters and the second is to see how well the corresponding models predict unseen time periods. We experiment with toy data and real data concerning the HIV epidemic in Cuba. 

\subsection{Toy Example}

Our simulated epidemic is generated using the following parameters: the number of initial infected individuals $|\mathcal{I}_0 | = 100$, probability of changing partner $\alpha = 0.9$, random detection rate $\gamma = 0.001$, contact tracing rate $\beta = 0.001$, sexual contact rate $\lambda = 0.1$, and infection probability $\sigma = 0.005$.  The graph contains $M = 5000$ individuals and we simulate the epidemic for 1000 days, making observations for the purposes of graph matching every 100 days. The initial graph has a hidden degree sequence generated using a power law with exponent $2$, there is an equal proportion of individuals from each gender and a proportion of $0.05$ individuals are labelled as bisexual.  To test the variability of the generated networks we simulated 10 epidemics under different random seeds and found a mean number of non-detected infected individuals of 39.8 (with standard deviation 6.6) and a mean number of detected individuals of 94.4 (with standard deviation 4.9). 

Next we attempt to recover the parameters of the model using ABC and the graph match objective $\Phi_{\pi}(\mathcal{G},\widehat{\mathcal{G}})$, with respect to initial simulated graph. We fix $\omega = 0.5$, $\varepsilon =0.5$ and the graph matching regularisation parameter $\nu = 0.2$ as we are more interested in the structure as opposed to matching the covariates of the vertices. The covariates for the vertices are: gender, sexual orientation, detection time and detection type (contact tracing or random detection). Graph matching is performed using the quadratic convex relaxation algorithm of \cite{zaslavskiy08path}. Note that we do not explicitly model differences in contact rates and infection probabilities between different covariate tuples as this would complicate the model and make parameter selection much more costly.  We modify the ABC procedure slightly by setting $\epsilon_1 = 0.8$ and then computing $\epsilon_i$, $i>1$, as the mean objective of the particles at population $i-1$. We stop ABC when all accepted particles have an objective value less than $\kappa = 0.3$ and take a population size of 50 particles.

To use ABC, we fix prior distributions and permutation kernels as follows. For the number of initial infected individuals $|\mathcal{I}_0 |$ we use the truncated discrete normal distribution as the prior with range $[0, 1500]$. The discrete normal distribution is used for the perturbation kernel. The prior distributions of $\alpha$ and $\sigma$ are truncated normal distributions with range $[0, 1]$ and their perturbation kernels are normal distributions. Finally, for the rate parameters the priors are gamma distributions and the corresponding perturbation kernels are normal distributions. The complete set of parameters for the stochastic model are written $\thetav = [|\mathcal{I}_0 | \; \alpha \; \gamma \; \beta \; \lambda \; \sigma]^T$ and we denote the mean and standard deviations of the prior distribution as $\mu_\pi(\thetav)$ and $\sigma_\phi(\thetav)$ respectively. We set $\mu_\pi(\thetav) = [100 \; 0.9 \; 0.001 \; 0.001 \; 0.1 \; 0.005]^T$ and $\sigma_\phi(\thetav) = \mu_\pi(\thetav)/10$. To evaluate the standard deviation of the perturbation kernels at iteration $i$, we use one fifth of the standard deviation of the particles at iteration $i-1$. At the end of the ABC process we use the parameters obtained for the epidemic models to resimulate them and record the number of infectives and detections. 

\begin{table}[ht] 
\centering 
\begin{tabular}{l | l l}
\hline
& Real & Estimate \\ 
\hline 
$|\mathcal{I}_0 |$ & 100.0000 (10.0000) &  99.2800 (4.2002)\\
$\alpha$ & 0.9000 (0.0900) &  0.8747 (0.0694)\\
$\gamma$ & 0.0010 (0.001) & 0.0011 (0.0002)\\
$\beta$ & 0.0010 (0.001) &  0.0010 (0.0003)\\
$\lambda$ & 0.1000 (0.100) & 0.0773 (0.0568)\\
$\sigma$ & 0.0050 (0.005)  & 0.0082 (0.0029)\\
\hline
\end{tabular} 
\caption{Real and estimated values using ABC correct to 4 d.p. with standard deviations in parenthesis.} 
\label{tab:abcSimulatedData} 
\end{table} 

\begin{figure}[!ht]
  \centering
\subfigure[Detections and infections]{\includegraphics[width=0.45\textwidth]{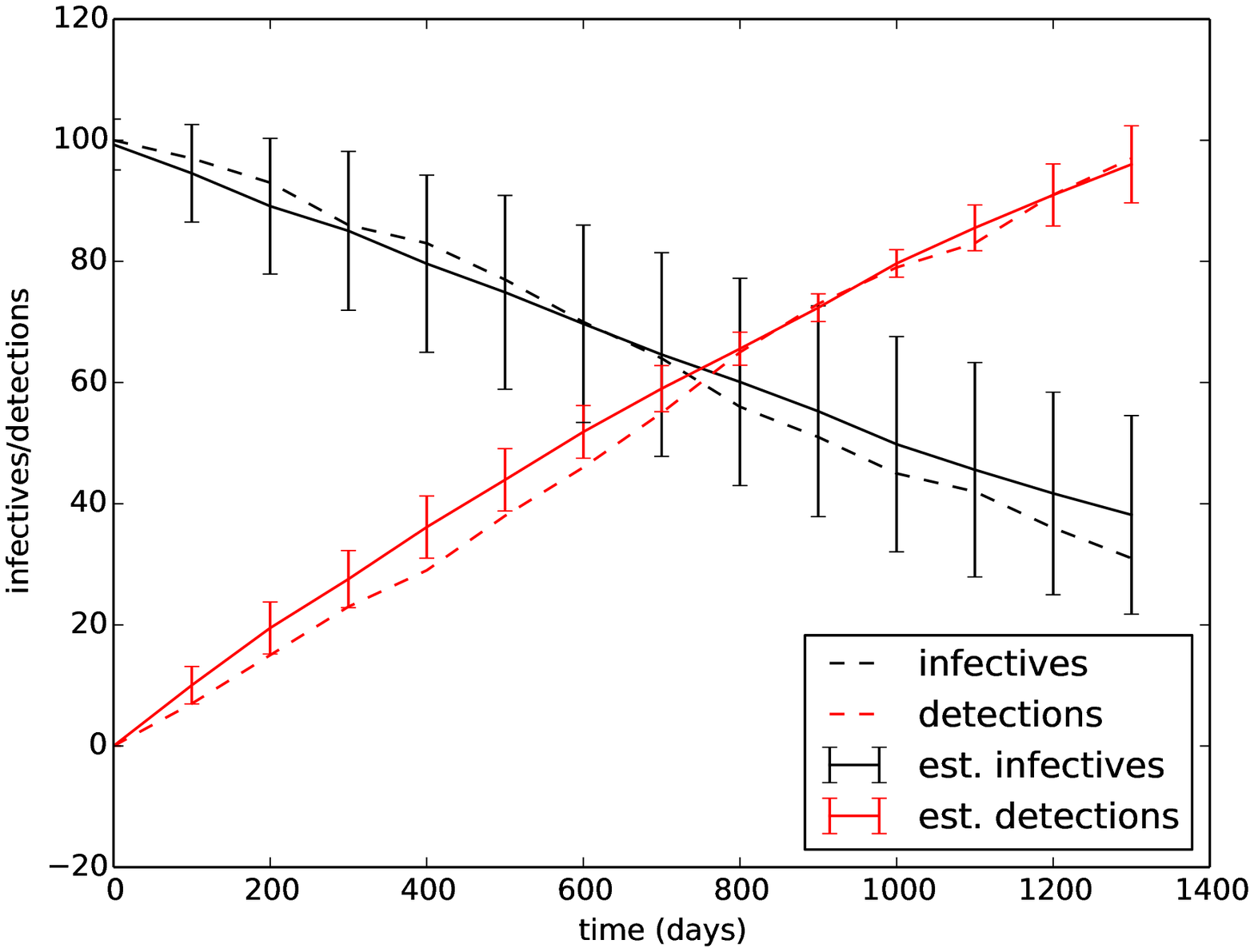}} 
\subfigure[Detection types]{\includegraphics[width=0.45\textwidth]{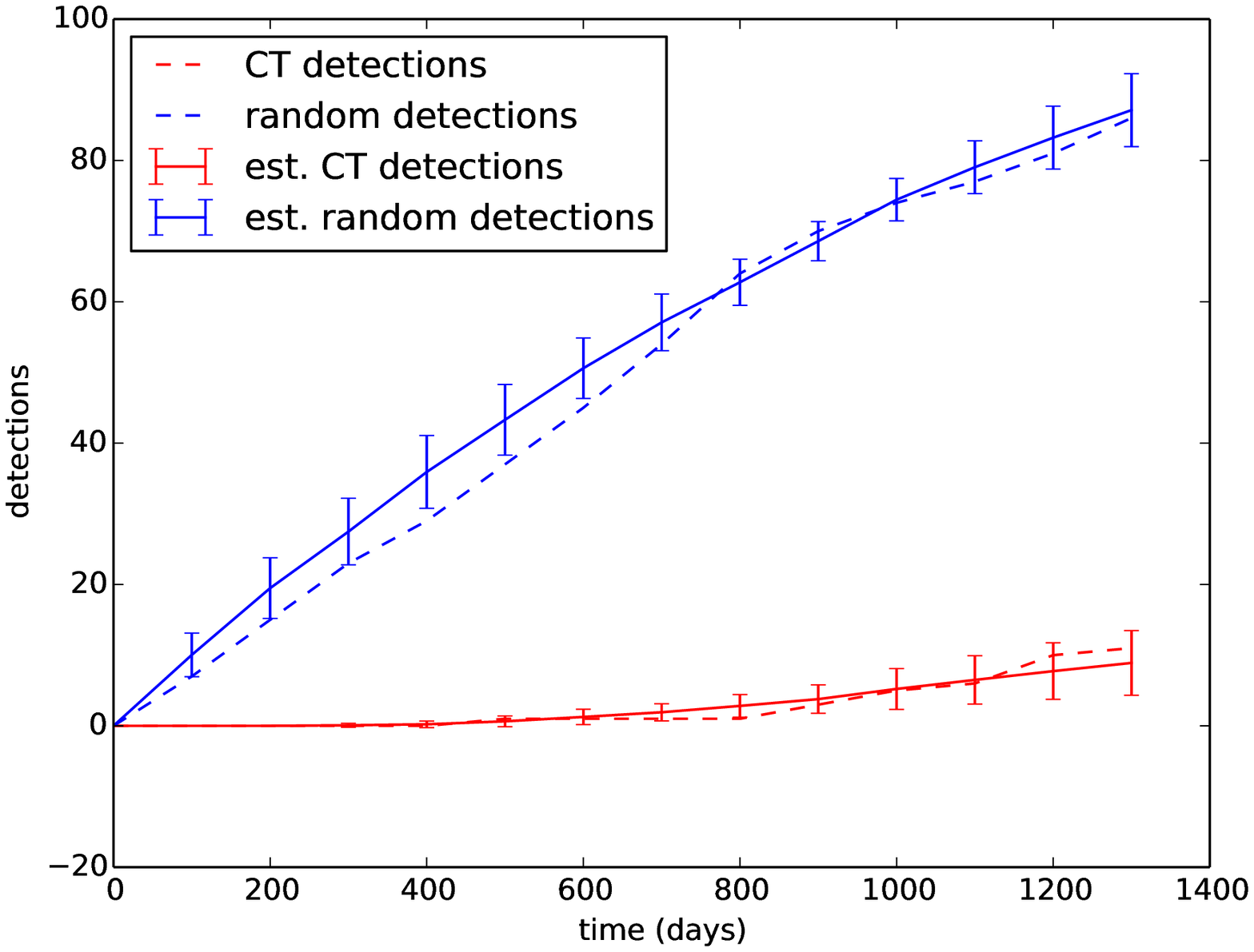}\label{fig:randCTToy}}
\caption{Detections for 50 simulated epidemics under the parameters learnt using ABC on the simulated data. Error bars represent the standard deviation.}\label{fig:toyMatch}
\end{figure}

The results for the ABC procedure are shown in Table \ref{tab:abcSimulatedData}, and we see that there is generally a close match for the real and estimated parameters with the exception of the contact rate and infection probability. Moreover, the standard deviations of the estimated parameters are significantly smaller than the equivalent deviations used for the corresponding prior distributions. We would not expect near-identical parameters due to the stochastic nature of the epidemic, which is apparent in terms of the infections and detections for the ABC-estimated models and the ``ideal'' one shown in Figure \ref{fig:toyMatch}. In this case a slightly higher infection rate and lower contact rate produces a similar number of detections to the ideal case particularly in the predicted period. The number of infectives on average match the ideal number well, however the standard deviation is large, and furthermore there appears to be an overestimation of infectives in the test period. In general we should not expect accurate prediction of the number of infectives since we match only on the detected vertices. Figure \ref{fig:randCTToy} shows the breakdown of the detections between those that are randomly detected and those that are contact traced. Again we see a tight correspondence between the real and estimated models although the estimated models tend to overestimate random detections slightly for the first 700 days. 

\subsection{Cuban HIV Epidemic} 

We next try to fit the model to the real epidemic data using ABC in a similar manner to that described above.  For training purposes we consider intervals starting and ending on the 1st of January: 1990-1992 1992-1994, 1998-1999 and 2002-2003. These intervals have been chosen according to the temporal behaviour of the real epidemic as described in \cite{dhanjal2011evolution}. In particular the first few years of the recorded epidemic after 1986 are avoided due to significant changes occurring during this period. We include the period 1990-1992 since it occurs just after a sudden increase in the detection of isolated vertices in 1990 and it is interesting to observe the error of the model over this change. In 1997 there is a sudden increase in detection rate, and thus we model two periods after this point. The first is in 1998 and the second is towards the end of the recorded period, however not the final year since not all detected individuals at the end of 2004 have been entered into the database. 

In each simulated epidemic the initial detected graph is based on that of the real epidemic, and we use total graph size of $M=26945$, 5 times larger than the observed epidemic at the end of the recorded period. For ABC, we sample 20 particles and compute graph objectives in steps of $T/5$ using $\epsilon_1 = 0.8$ with successive $\epsilon$'s generated from the last as before. The ABC procedure stops when $\epsilon_i$ is smaller than $\kappa = 0.3$. The same prior and perturbation kernels as those from the toy experiment are used for the parameters. In this case we fix $\mu_\pi(\thetav) = [500 \; 0.5 \; 0.01 \; 0.1 \; 0.1 \; 0.1]^T$ and $\sigma_\pi(\thetav) = [400 \; 0.5 \; 0.01 \; 0.1 \; 0.1\; 0.1]^T$ and hence the prior is relatively uninformative. As before the standard deviation of each perturbation kernel at iteration $i$ is one fifth of the standard deviation of the particles at iteration $i-1$. After the learning stage we simulate using all accepted particles for a time of $1.2T$ to observe if the model is predictive for this period, recording graph properties at steps of $T/5$.  

\begin{figure}[!ht]
\centering
\subfigure[1990-1992]{\includegraphics[width=0.45\textwidth]{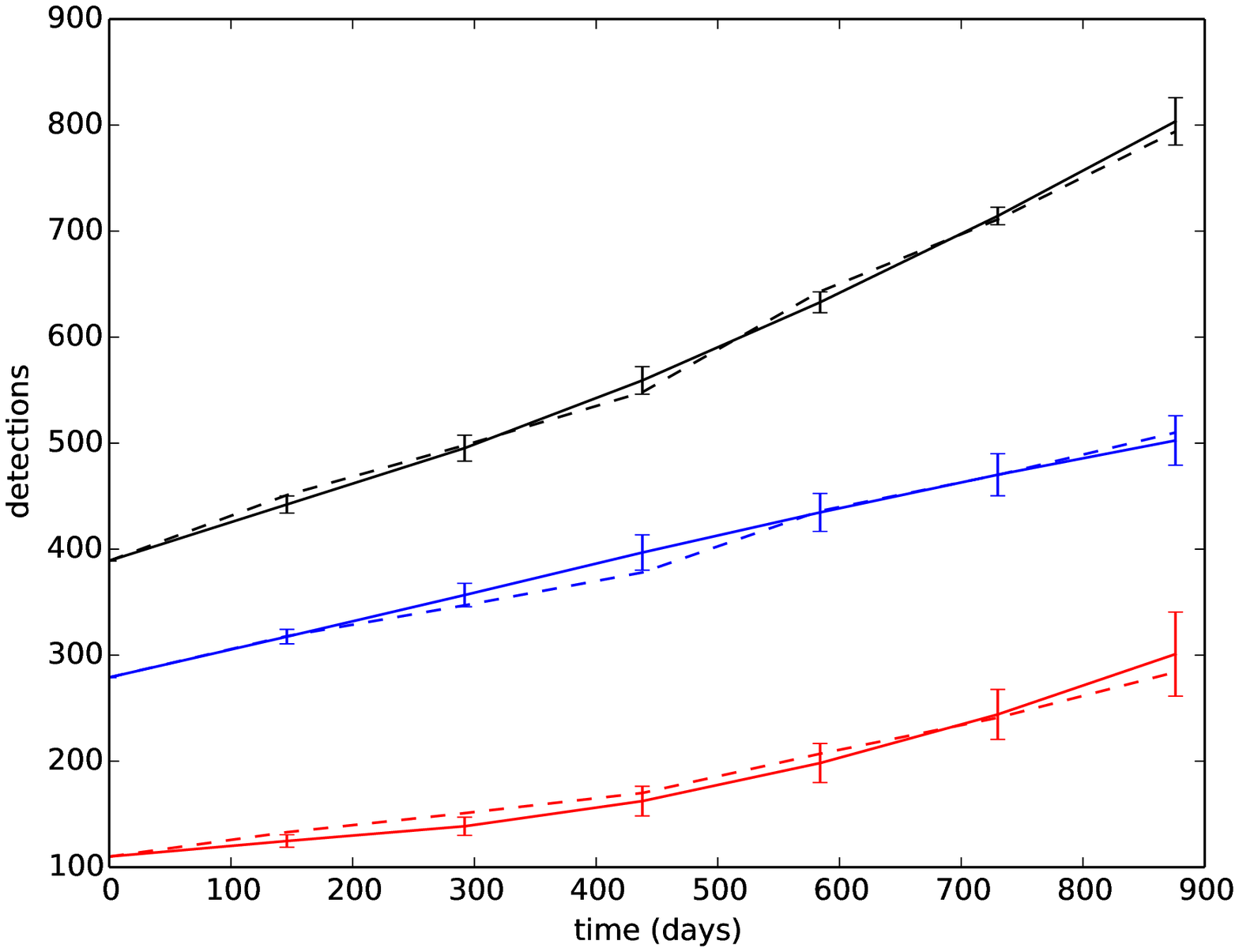}\label{subfig:realABCMatch0}}
\subfigure[1992-1994]{\includegraphics[width=0.45\textwidth]{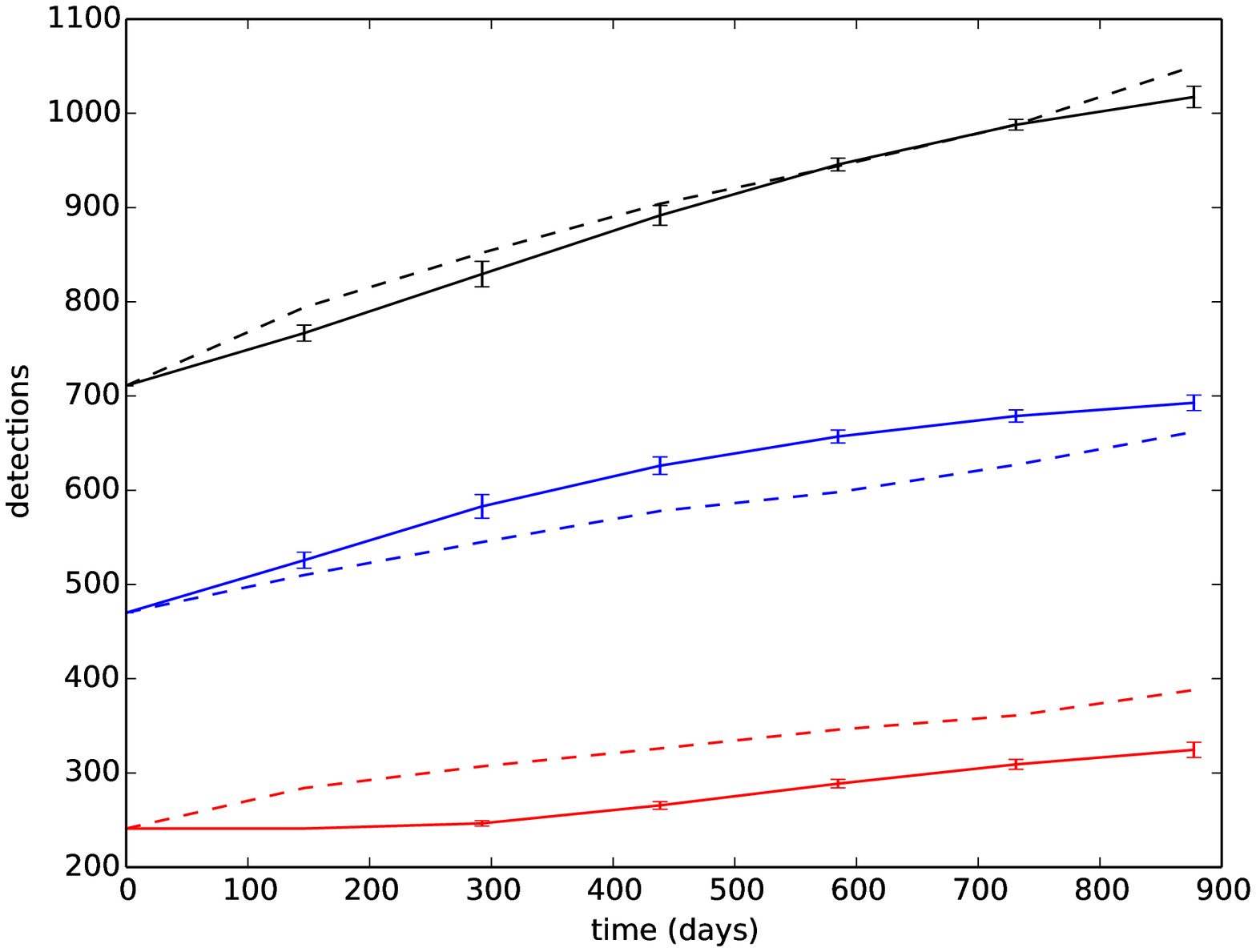}\label{subfig:realABCMatch1}}
\subfigure[1998-1999]{\includegraphics[width=0.45\textwidth]{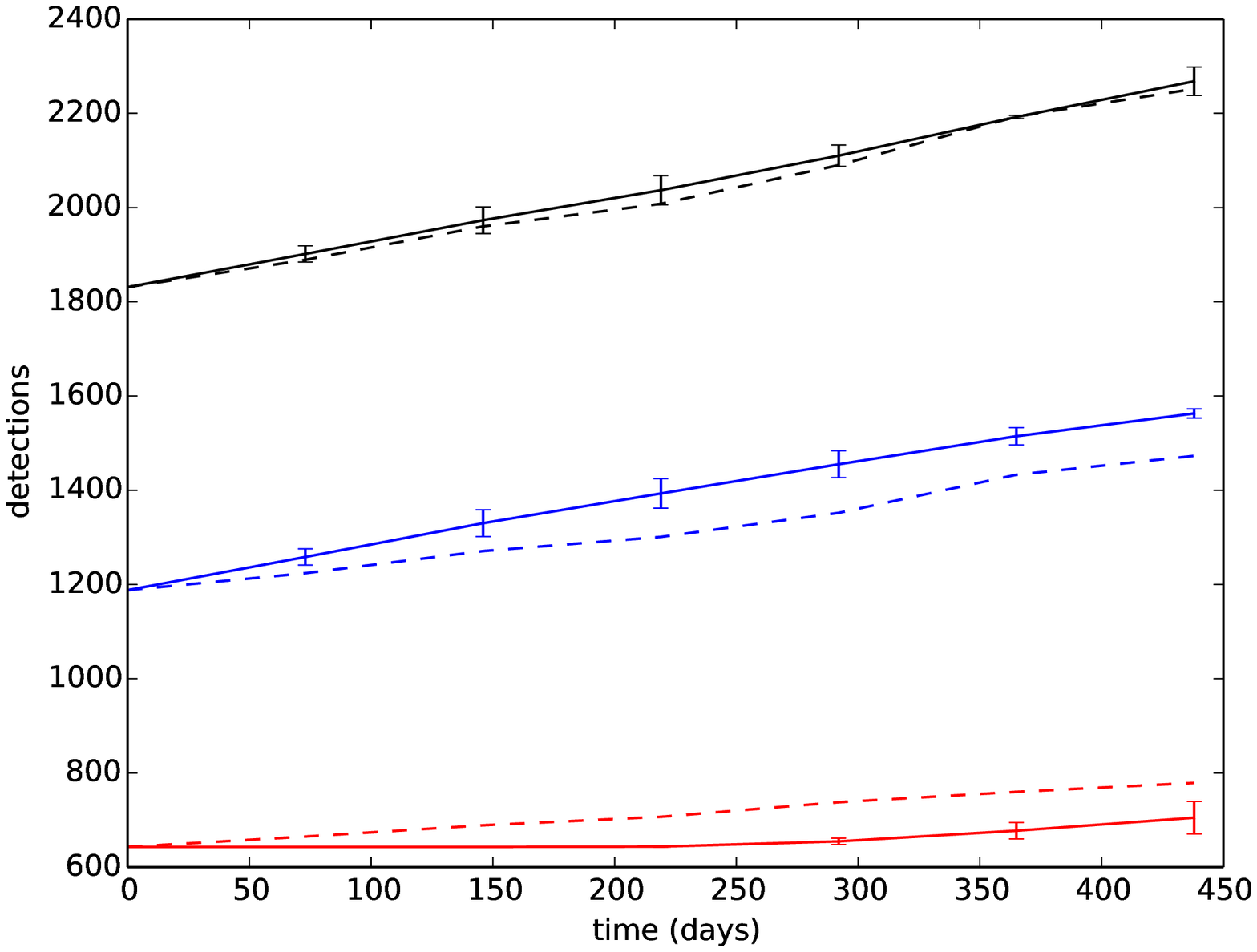}\label{subfig:realABCMatch2}}
\subfigure[2002-2003]{\includegraphics[width=0.45\textwidth]{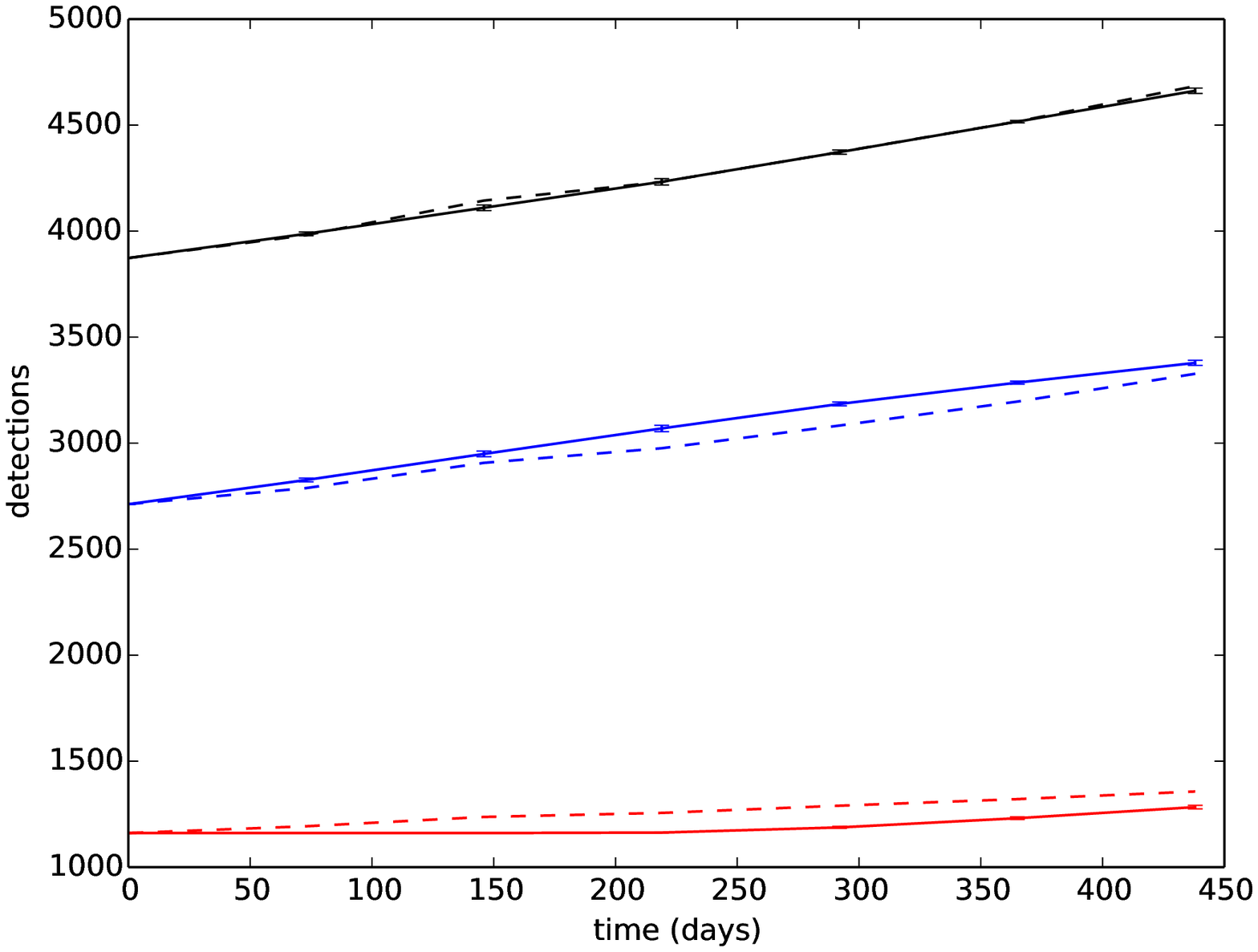}\label{subfig:realABCMatch3}}
\caption{Result of the ABC procedure to find parameters for the epidemic model on the real data. In black is the total number of detections, blue is the number of random detections and red are contact traced detections. The solid lines represent estimated figures where error bars show the standard deviation, and dashed lines represent the real figures.}\label{fig:realABCMatch}
\end{figure}

\begin{table*}[!ht] 
\centering 
\footnotesize 
\begin{tabular}{l | l  l | l l | l l | l l}
\hline
&   \multicolumn{2}{c}{1990-1992} &  \multicolumn{2}{c}{1992-1994}  &  \multicolumn{2}{c}{1998-1999} & \multicolumn{2}{c}{2002-2003}\\ 
&   $t_5$ &  $t_6$ &  $t_5$ &  $t_6$ &  $t_5$ &  $t_6$ &  $t_5$ &  $t_6$ \\ 
\hline
$|\mathcal{R}|$. & 711 & 794 & 988 & 1050 & 2193 & 2252 & 4517 & 4684\\
$|\mathcal{R}|$. est. & 714 (8) & 804 (22) & 988 (6) & 1017 (11) & 2192 (3) & 2268 (30) & 4516 (5) & 4662 (13)\\
\hline
RD & 470 & 510 & 627 & 662 & 1433 & 1473 & 3196 & 3327\\
RD est. & 470 (20) & 502 (23) & 679 (6) & 693 (8) & 1515 (18) & 1563 (10) & 3285 (7) & 3379 (12)\\
\hline
CT & 241 & 284 & 361 & 388 & 760 & 779 & 1321 & 1357\\
CT est. & 244 (24) & 301 (40) & 309 (5) & 325 (8) & 677 (17) & 705 (35) & 1231 (6) & 1283 (9)\\
\hline
LC & 301 & 359 & 457 & 483 & 1133 & 1158 & 2129 & 2174\\
LC est. & 153 (21) & 178 (40) & 347 (20) & 365 (28) & 1066 (81) & 1164 (105) & 2479 (33) & 2771 (49)\\
\hline
NC & 290 & 310 & 365 & 389 & 747 & 769 & 1659 & 1740\\
NC est. & 359 (23) & 375 (30) & 405 (9) & 402 (9) & 778 (38) & 762 (47) & 1488 (12) & 1422 (17)\\
\hline
$|E|$ & 571 & 679 & 925 & 970 & 1962 & 2007 & 3631 & 3729\\
$|E|$ est. & 393 (28) & 467 (49) & 735 (11) & 769 (18) & 1894 (56) & 2006 (80) & 3878 (28) & 4220 (49)\\
\hline
$\phi(\mathcal{G}_{t_i}, \widehat{\mathcal{G}}_{t_i})$ & .35 (.07) & .44 (.05) & .37 (.06) & 0.49 (.02) & .25 (.13) & .52 (.02) & .34 (.10) & .51 (.03)\\
\hline
\end{tabular} 
\caption{Some properties of the evolving graphs with standard deviations in parentheses. LC means largest connected component, and NC is the number of components.} 
\label{tab:realProps} 
\end{table*} 

Figure \ref{fig:realABCMatch} compares the number of real and estimated detections. We see that on average the estimated number of total detections are close to the real ones for the training periods. Furthermore, the predictions for the number of detections are accurate for the test time point, with the exception of the period 1992-1994. Here the estimated models underestimate the detections by just 33 individuals and most of these errors come from an underestimation in the number of contact traced detections, which in part are cancelled out by an overestimation in the number of random detections. It is worth noting that the number of detections could be estimated to a similar accuracy using curve fitting for example however in our case we can determine much more detailed properties of the simulated epidemic as we have complete networks.

Table \ref{tab:realProps} shows some further properties of the real and estimated graphs for time points $t_5$ and $t_6$. There is a tendency to underestimate the number of edges in the graphs for the periods 1990-1992 and 2002-2003. As one might expect this results in smaller largest components and fewer components overall when compared to the real epidemic. In 1998-1999 the corresponding figures are relatively accurate with an error in the number of vertices in the largest component of just 6 and the corresponding number of estimated components is 762 versus 769 in the real network. In general, we should not expect results too close to the real epidemic due to its complexity relative to the model we propose, for example immigration/emigration, changes in health policy and infection spread through Men having Sex with Men (MSM) contact. We also looked at the values of $\phi(\mathcal{G}_{t_i}, \widehat{\mathcal{G}}_{t_i})$ and we see that the graphs at $t_5$ are quite accurate with $\phi(\mathcal{G}_{t_5}, \widehat{\mathcal{G}}_{t_5}) \leq 0.37$. This increases on the test time point $t_6$, most significantly with 1998-1999 and 2002-2003. 

\begin{table}[!ht] 
\centering 
\begin{tabular}{l | l l l l}
\hline
& 1990-1992 & 1992-1994 & 1998-1999 & 2002-2003\\ 
\hline
$|\mathcal{I}_{t_0}|$ & 537.8500 (50.9228) & 158.2000 (5.1342) & 359.2500 (81.1664) & 421.8500 (5.4889)\\
$\alpha$ & 0.7766 (0.0352) & 0.8200 (0.0234) & 0.4309 (0.1669) & 0.4718 (0.0124)\\
$\gamma$ & 0.0005 (0.0001) & 0.0024 (0.0001) & 0.0031 (0.0015) & 0.0037 (0.0001)\\
$\beta$ & 0.0913 (0.0316) & 0.0342 (0.0044) & 0.0373 (0.0375) & 0.0288 (0.0038)\\
$\lambda$ & 0.0187 (0.0029) & 0.0315 (0.0012) & 0.0281 (0.0053) & 0.0360 (0.0005)\\
$\sigma$ & 0.0604 (0.0166) & 0.1245 (0.0058) & 0.1408 (0.0286) & 0.1828 (0.0022)\\
\hline
$\hat{\alpha}$  & 0.0005 & 0.0017 & 0.0015 & 0.0024 \\ 
\hline
\end{tabular} 
\caption{Estimated values for real epidemic correct to 4 d.p. with standard deviations in parenthesis. The final line, $\hat{\alpha}$, represents the estimated random detection rate from the start of each period using the real epidemic.} 
\label{tab:realParams} 
\end{table} 

Some additional insight into the generated epidemics can be garnered from Table \ref{tab:realParams} which shows estimated parameter values. Note firstly that the random detection rate increases over time, and this correlates with the trend observed in the real epidemic. As a comparison we compute the random detection rate per infected person for the real epidemic based on the start of each time period, $t_1 - t_0$, denoted $\hat{\alpha}$ in the table. This value is computed by $(|\mathcal{R}^r_{t_1}| - |\mathcal{R}^r_{t_0}|)/((t_1 - t_0)|\mathcal{I}_{t_0}|)$ where $\mathcal{R}^r_i$ is the set of individuals detected randomly at time $i$ in the real data and $|\mathcal{I}_{t_0}|$ is the initial infected individuals in the simulated epidemic. There is a general increase in random detection rate over time. We might expect that these values underestimate the ones found in the simulation as random detections are overestimated slightly in practice. The high contact tracing rate in 1990-1992 corresponds exactly to the description of the epidemic given in \cite{dhanjal2011evolution}. Specifically after 1989 the number of individuals detected using contact tracing increases however the rate of random detections also increase at this point, and outpaces that of contact tracing.  The increase in infection rate over time can be explained by the discovery of infected individuals in new geographical locations in the epidemic, for example during 1990-1992 there is a heightened detection rate in Pinar del Rio. As explained in \cite{dhanjal2011evolution}, one reason for the increase in the number of detections in 1997 is the lack of detections in the beginning of the 1990's which created a pool of undetected individuals. In the estimated parameters for the stochastic model this is approximated using a high infection rate $\sigma$ and low value of $\alpha$. 

We conclude the analysis of the generated networks by making some remarks on the computation time, measured on an Intel Xeon E5430 running at 2.66GHz with 32GB RAM. The computation of each model is relatively efficient taking just 326 seconds on average for the final iteration of ABC and for the most costly time period, 2002-2003. Of this, the majority of the time is spent on graph matching, specifically 281 seconds. An average of 45 seconds was required to simulate an increase of 793 vertices in the detection graph and approximately 6000 unique contacts. 

\section{Conclusions}

In this paper, we introduced a novel graph-based SIR epidemic model which did not specify the graph structure in advance. The resulting model simulates events of contact, infection and detection using Poisson processes, is efficient to compute and general enough to encapsulate a wide variety of epidemics. Here we specified a model for the spread of HIV, with an application to the Cuban epidemic. The graph generated from the model was matched to a partially observed real epidemic graph using state-of-the-art graph matching in conjunction with ABC for parameter estimation. We saw a close fit of the number of estimated detections as well as several graph properties of the real epidemic. Furthermore, several trends in the true epidemic were shown to be explained by the estimated parameters of the stochastic model. 

With this work one can naturally adapt the model proposed to other epidemics where data is available. Furthermore, one can attempt to improve the efficiency of large graph matching by looking at online methods which can leverage previous matches in time evolving graphs. This would enable not just the study of large graphs but also more complicated stochastic models which may better fit the dynamics of real epidemics. 

\section*{Acknowledgements} The authors are grateful to Dr. H. De Arazoza of the University of La Havana and to Dr. J. Perez of the National Institute of Tropical Diseases in Cuba for granting access to the HIV-AIDS database. We are also grateful to Viet Chi Tran for discussions concerning the epidemic modelling. This research was financed by the ANR Viroscopy (ANR-08-SYSC-016-03).

\bibliographystyle{abbrv}
\bibliography{references,my-references}

 \appendix

\section{Simulation Algorithm}\label{sec:alg}

The pseudo code for the stochastic epidemic model is shown in Algorithm \ref{alg:epidemicModel}. Note that the sets of individuals in the states Susceptible, Infected and Removed are denoted at time $t\in \mathbb{R}_+$ by $\mathcal{S}_t$, $\mathcal{I}_t$ and $\mathcal{R}_t$. We denote by $|\mathcal{S}_t|$, $|\mathcal{I}_t|$ and $|\mathcal{R}_t|$ the sizes of these classes and the population is assumed to be closed, with size $M$. The algorithm is initialised with rate functions for contact and detection, $c: \mathcal{X} \times \mathcal{X} \rightarrow \mathbb{R}^+$ and $g: \mathcal{X} \rightarrow \mathbb{R}^+$ which indicate how frequent each respective event occurs based on the characteristics of vertices at time $t$. One also supplies a probability $f: \mathcal{X} \times \mathcal{X} \rightarrow [0, 1]$ of an infection occurring between two individuals who have contact.  An initial contact graph $G_0$ is required, and in order for the simulation to give non-trivial results there must be at least one infected individual. The initial SIR states of individuals are given in the sets $\mathcal{S}_0$, $\mathcal{I}_0$, $\mathcal{R}_0$ and also for simplicity mirrored in the vertex labels $\Xm_0$. 

\begin{algorithm}
\caption{Modelling an epidemic spread over a network}
\label{alg:epidemicModel}
\begin{algorithmic}[1]
\STATE Input: $t=0$, $T > 0$,  $G_0 = (V, E_0, \Xm_0)$ with $|V|=M$,  $\mathcal{S}_0$, $\mathcal{I}_0$, $\mathcal{R}_0$, contact rate function $c$, infection probability $f$, detection rate $g$
\STATE $\mathcal{T} = \{t\}$ 
\WHILE{$(t < T \mbox{ and } |\mathcal{I}_t| \neq 0)$} \label{lin:while}
\STATE Sample exponential variable $\tau$ with rate $\hat{\rho}_t = \sum_{i \not= j} \hat{\Am}_{ij}^t + \sum_{i=1}^M \hat{\bv}_{i}^t$ where $\hat{\Am}^t$ and $\hat{\bv}^t$ are upper bounds on contact and detection rates for $s>t$ \label{lin:rho}
\STATE Let $t \leftarrow t+\tau$, and update $\xv_i^t$ for all $i$ \label{lin:updateTime}
\STATE Contact rates $\Am^t \in \mathbb{R}^{M\times M}$, $\Am^t_{ij} = c(\xv_i^t, \xv_j^t)$ for $(v_i^t, v_j^t) \in ((\mathcal{S}_t \cup \mathcal{I}_t) \times \mathcal{I}_t$) \label{lin:contactRates}
\STATE Detection rates $\bv^t \in \mathbb{R}^{M}$, $\bv^t_i = g(\xv_i^t)$ for $v_i^t \in \mathcal{I}_t$ \label{lin:detectRates}
\STATE Choose event $\beta_t$ according to the probabilities $\Am_{ij}^t/\widehat{\rho}_t$  and $\bv_i^t/\widehat{\rho}_t$  \label{lin:chooseEvent}
\IF{$\beta_t$ is contact event $(v_i, v_j)$ and $v_i \in \mathcal{I}_t$}
\STATE  $E_t \leftarrow E_t \cup \{(v_i, v_j)\}$ 
\IF{infection occurs with probability $f(\xv_i^t, \xv_j^t)$ } \label{lin:infectProb}
\STATE $\mathcal{I}_t \leftarrow \mathcal{I}_t \cup \{v_j\}$
\ENDIF
\ELSIF{$\beta_t$ detection event of $v_i$}
\STATE $\mathcal{R}_t \leftarrow \mathcal{R}_t \cup \{v_i\}$
\ENDIF
\STATE Update $\mathcal{T} \leftarrow \mathcal{T} \cup \{t\}$
\ENDWHILE
\STATE Output: $\mathcal{S}_t$, $\mathcal{I}_t$, $\mathcal{R}_t, G_t$, $\forall t \in \mathcal{T}$.
\end{algorithmic}
\end{algorithm}

After initialising variables, the loop starts by upper bounding the contact and detection rates in Step \ref{lin:rho} with $\hat{\rho}_t$. Notice that this upper bound depends only on the composition of the population at time $t$, and allows us to decide when the next event occurs by sampling from a Poisson distribution with this rate. Steps \ref{lin:contactRates} and \ref{lin:detectRates} compute rates for contact and detection events for all of the vertices in the graph and store the rates at time $t$ in $\Am^t \in \mathbb{R}^{M \times M}$ and $\bv^t \in \mathbb{R}^{M}$ respectively. Notice that $\Am^t$ must be symmetric and is typically a sparse matrix. The computation of event probabilities is computed from the rates, specifically the probability of each event is proportional to its rate. It follows that one chooses an event $\beta_t$ according to the probabilities $\Am_{ij}^t/\widehat{\rho}_t$ (individual $v_i$ has a contact with $v_j$) and $\bv_i^t/\widehat{\rho}_t$ (individual $v_i$ is removed). With a probability $(\widehat{\rho}_t-\sum_{i\not= j}\Am_{ij}^t-\sum_{i=1}^M \bv_i^t)/\widehat{\rho}_t$, nothing happens. At Step \ref{lin:infectProb} we check if a contact results in an infection based on the infection probability function $f$, and then the loop iterates after updating variables accordingly.

\end{document}

%% file: CsdMacros.tex
\newcommand{\bv}{\textbf{b}}

\newcommand{\xv}{\textbf{x}}

\newcommand{\thetav}{\mbox{\boldmath$\theta$}}

\newcommand{\Am}{\textbf{A}}

\newcommand{\Cm}{\textbf{C}}

\newcommand{\Pm}{\textbf{P}}

\newcommand{\Um}{\textbf{U}}
\newcommand{\Vm}{\textbf{V}}

\newcommand{\Xm}{\textbf{X}}
